\documentclass[conference]{IEEEtran}
\usepackage{amsmath,amsfonts}
\usepackage{amsthm}
\usepackage{cases,amssymb}
\usepackage[caption=false,font=normalsize,labelfont=sf,textfont=sf]{subfig}
\usepackage{array}
\usepackage[linesnumbered,ruled,vlined]{algorithm2e}
\usepackage{cite,setspace,bm,enumerate}
\usepackage{color}
\usepackage{stfloats}
\usepackage{caption2}  
\usepackage{textcomp}
\usepackage{url}
\usepackage{verbatim}
\usepackage{graphicx}

\def\BibTeX{{\rm B\kern-.05em{\sc i\kern-.025em b}\kern-.08em
    T\kern-.1667em\lower.7ex\hbox{E}\kern-.125emX}}
\usepackage{balance}

\DeclareMathOperator*{\argmax}{arg\,max}

\theoremstyle{definition}

\begin{document}

\title{A Three-Party Repeated Coalition Formation Game for PLS in Wireless Communications with IRSs}

\author{ \small Haipeng Zhou$^{\ast}$,  Ruoyang Chen$^{\ast}$, Changyan Yi$^{\ast}$, Juan Li$^{\ast}$ and Jun Cai$^{\dagger}$\\
\IEEEauthorblockA{\text{\small $^{\ast}$College of Computer Science and Technology, Nanjing University of Aeronautics and Astronautics, Nanjing, China} \\
\small $^{\dagger}$Department of Electrical and Computer Engineering, Concordia University, Montr\'{e}al, QC, H3G 1M8, Canada \\
\text{\small Email: \{haipengzhou, ruoyangchen, changyan.yi, juanli\}@nuaa.edu.cn,  jun.cai@concordia.ca}\\
}
}

\maketitle

\begin{abstract}
    In this paper, a repeated coalition formation game (RCFG) with dynamic decision-making for physical layer security (PLS) in wireless communications with intelligent reflecting surfaces (IRSs) has been investigated. In the considered system, one central legitimate transmitter (LT) aims to transmit secret signals to a group of legitimate receivers (LRs) under the threat of a proactive eavesdropper (EV), while there exist a number of third-party IRSs (TIRSs) which can choose to form a coalition with either legitimate pairs (LPs) or the EV to improve their respective performances in exchange for potential benefits (e.g., payments). Unlike existing works that commonly restricted to friendly IRSs or malicious IRSs only, we study the complicated dynamic ally-adversary relationships among LPs, EV and TIRSs, under unpredictable wireless channel conditions, and introduce a RCFG to model their long-term strategic interactions. Particularly, we first analyze the existence of Nash equilibrium (NE) in the formulated RCFG, and then propose a switch operations-based coalition selection along with a deep reinforcement learning (DRL)-based algorithm for obtaining such equilibrium. Simulations examine the feasibility of the proposed algorithm and show its superiority over counterparts.
\end{abstract}


\section{INTRODUCTION}
Due to the open broadcast nature of wireless channels, wireless signals are vulnerable to eavesdropping. This motivates the investigation on physical layer security (PLS) which exploits the intrinsic physical properties of wireless channels to against the eavesdropper (EV). Compared with traditional encryption/decryption-based methods implemented in higher layers of open system interconnection (OSI) model, PLS exhibits extraordinary advantages in low computational complexity and resource consumption, and thus has been widely employed in a variety of applications\cite{31}.

As a promising technology to enable programmable wireless environment, intelligent reflecting surfaces (IRSs), has recently attracted great attentions, due to its capability of adjusting the phase shifts of its passive reflecting elements to reconfigure the wireless channels\cite{4}. Despite that PLS-aware wireless communications with IRSs has been extensively studied in existing works\cite{9,5}, most of them did not fully explore the relationships among all network participants, including legitimate pairs (LPs), EV, and IRSs under their potential selfishnesses, especially
ignoring the impact brought by the existence of third-party IRSs (TIRSs). Particularly, since TIRSs do not share the same interest with either LPs or EV, they may be selfishly pursuing their own interests by assisting either LPs or EV depending on different situations, i.e., i) when TIRSs can acquire more benefit from LPs than that from EV, they may form a coalition with LPs by adjusting their elements' phase shifts to increase LPs' secrecy rates; and ii) when TIRSs can acquire more benefit from EV than that from LPs, they may form a coalition with EV by adjusting their elements' phase shifts to boost EV's eavesdropping rates \cite{10279584}. Obviously, such ally-adversary relationships among these three parties (i.e., LPs, EV and TIRSs) may not be predefined, while it is worthy to be carefully analyzed. However, to the best of our knowledge, this crucial issue has not yet been studied in the literature, and is very challenging as outlined below.
\begin{itemize}
     \item Apart from potential coalition formations among LPs, EV and TIRSs, to enhance each party's performance in PLS, LPs need to determine their transmit beamforming vectors, EV should determine its jamming beamforming vectors, and TIRSs are required to optimize their phase-shifting matrices. All these decisions are multi-dimensional with complex features, and may be inherently interdependent to each other, resulting in a tightly coupled decision-making process. Moreover, such process in turn influences the coalition formation among three parties, necessitating a coalition formation game with multi-dimensional strategies to model and analyze the formation of the coalition structure (i.e., coalition partitions).
     \item Owning the system uncertainties in wireless communications, e.g., dynamic channel conditions, the strategies of LPs, EV and TIRSs may be dynamically adjusted, leading to the dynamic evolution in coalition structures among them. This motivates a dynamic transition on coalition formation game to a repeated coalition formation game (RCFG), in which any two of the three parties may temporarily form coalitions (becoming allies) and dynamically evolve in different time under system dynamics.
\end{itemize}

To overcome the above challenges, in this paper, we propose a RCFG framework to model the long-term coalition formations with dynamic decision-making among LPs, EV and TIRSs. To be more specific, we first model the utility functions for all three parties in PLS, and then formulate the long-term optimization problem for respectively maximizing the cumulative utility of each party, considering their dynamic-evolved strategies. A three-party RCFG is consequently developed for analyzing the dynamic relationships among them. Taking into account i) the hedonic nature of the three parties' coalition formation, switch operations are employed for each party's dynamic optimal coalition selection; and ii) the Markov properties of decision-making within a coalition, a deep reinforcement learning (DRL) based algorithm is proposed for obtaining the optimal long-term strategies.

The main contributions of this paper are summarized as follows:
\begin{itemize}
    \item To model the long-term coalition formations with dynamic decision-making among LPs, EV and TIRSs, a three-party RCFG is rigorously formulated and analyzed.
    \item To optimize the three parties' respective long-term performances under their dynamic coalition formations, switch operations-based coalition selections along with a DRL-based algorithm are developed for obtaining the equilibrium of the proposed RCFG.
    \item Simulations examine the feasibility of the proposed algorithm and show its superiority over counterparts.
\end{itemize}

The rest of this paper is organized as follows. Section II introduces the system model and problem formulation. In Section III, the RCFG is formulated and analyzed, and then a switch operations-based coalition selection along with a DRL-based algorithm is proposed. Simulation results are given in Section IV, followed by the conclusion in Section V.

\section{System Model and Problem Formulation}

\subsection{System Model}
\begin{figure}[!t]
	\centering
	\includegraphics[width=0.65\columnwidth]{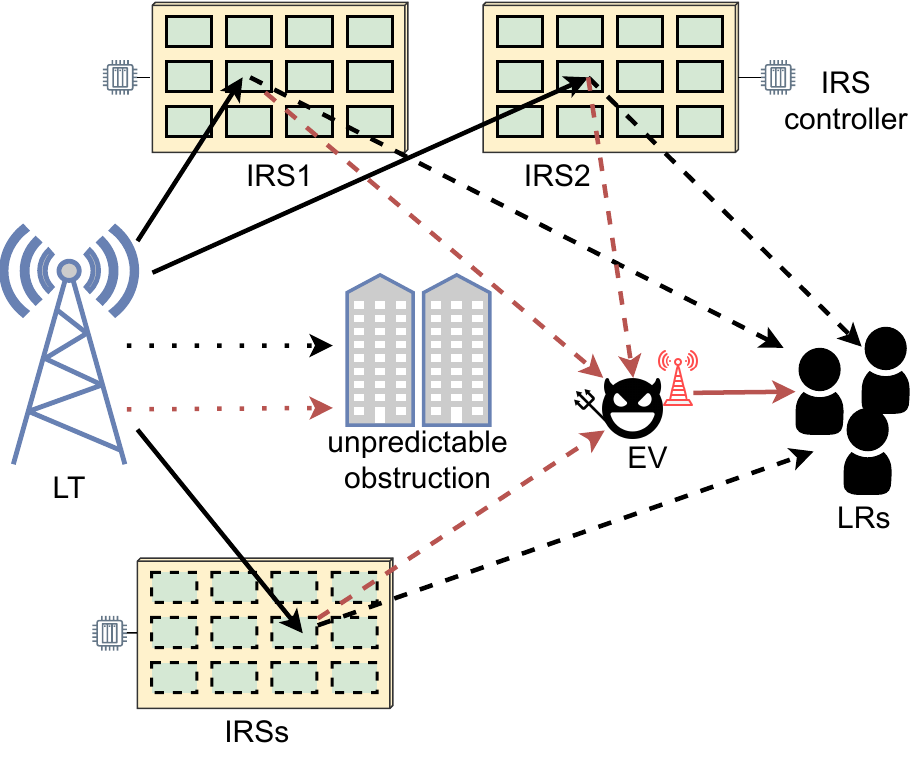}
	\caption{PLS-aware wireless communication system with IRSs.}\label{pic:model}
	\vspace{-1.6em}
\end{figure}
As shown in Fig. \ref{pic:model}, we consider a downlink wireless communication system  consists of one central legitimate transmitter (LT), which can be base station or access point in practice, $K$ IRSs, denoted by the set $\mathcal{K}=\{1,...,K\}$, $L$ legitimate receivers (LRs), denoted by the set $\mathcal{L}=\{1,...,L\}$, and one EV. To depict time-varying uncertainties of the wireless system, a time-slotted operation framework is studied, where $t \in \{1,...,T\}$ denotes the index of time slots. The LT with $M_a$ transmitting antennas send $L$ independent confidential signals with one stream to the LRs with one receiving antenna each over the same frequency band, simultaneously. In each time slot, $K$ IRSs with $N$ passive reflecting elements each are able to reflect the incident signals by dynamically adjusting the reflection amplitude and/or phase shift of each element under the control of a micro-controller\cite{5}. The EV is equipped with one antenna for eavesdropping and $M_e$ antennas for jamming. 

We consider quasi-static block-fading channels and all channels are assumed to remain approximately constant within each time slot. The signals reflected by IRS two or more times are ignored due to the severe distance-product power loss over multiple reflections\cite{19}. We use $G^H_{ak,t} \in \mathbb{C}^{N\times M_a} , h^H_{ai,t} \in \mathbb{C}^{1 \times M_a} , h^H_{ae,t} \in \mathbb{C}^{1\times M_a}, g^H_{ki,t} \in \mathbb{C}^{1 \times N} , g^H_{ke,t} \in \mathbb{C}^{1 \times N}, h^H_{ei,t} \in \mathbb{C}^{1 \times M_e} , h^H_{ee,t}\in \mathbb{C}^{1\times M_e}$ to denote the baseband equivalent channel from LT to $k$th IRS, LT to $i$th LR, LT to EV, $k$th IRS to $i$th LR, $k$th IRS to EV, EV to $i$th LR and EV's self-interference (SI), respectively, in which $\mathbb{C}^{m \times n}$ denotes the set of $m \times n$ complex matrix. The LT and EV employ linear transmit precoding\cite{4}, and the confidential signal transmitted from LT to $i$th LR can be described as $s_{i,t}=w_i(t)d_{i,t}, i \in \mathcal{L},$ where {\small$w_i(t) \in \mathbb{C}^{1\times M_a}$} represents the transmit beamforming vector of $i$th LR, and $d_i$ denotes the transmitted data.

Both LPs and EV can be assisted by multiple TIRSs to receive reconfigured signals, especially in scenarios where no Line-of-Sight (NLoS) channel exists from LT to LRs and EV due to unpredictable obstructions. The diagonal phase-shifting matrix at $k$th IRS in time slot $t$ is denoted as $\Phi_k(t)=diag(A_{k,1}e^{j\theta_{k,1}(t)},...,A_{k,N}e^{j\theta_{k,N}(t)}) \in \mathbb{C}^{N\times N}$, in which $A_{k,n} \in [0,1]$ represents the amplitude reflection coefficient, and $\theta_{k,n}(t) \in [0,2\pi]$ stands for the phase shift of $n$th element on $k$th IRS. As each phase shift is desired to be desired to achieve full reflection, we consider that $A_{k,n} = 1$ in \cite{5}. The received signal at $i$th LR can be expressed as
\begin{equation}
\begin{split}
 y_{i,t} = &(\sum\nolimits _{k=1}^K g^H_{ki,t}{\Phi }_k(t)G^H_{ak,t}+h^H_{ai,t})w_i(t)s_{i,t}+ \\&\sum \nolimits _{j\ne i}(\sum \nolimits _{k=1}^K g^H_{ki,t}{\Phi}_k(t)G^H_{ak,t}+h^H_{ai,t})w_j(t)s_{j,t}\\&+\sum \nolimits _{j=1}^L h^H_{ei,t}f_j(t)z_{j,t}+n_0,
\end{split}
\end{equation}
where $n_0 \sim \mathcal{CN}(0,\sigma^2_0)$ represents the complex Gaussian distribution with zero mean and variance $\sigma^2_0$ represents the complex additive Gaussian white noise (AWGN) at $i$th LR, of which $\mathcal{CN}(m,n)$ is the complex Gaussian distribution with mean $m$ and variance $n$, $z_{i,t} \sim \mathcal{CN}(0,1)$ denotes the jamming signal from EV, and $f_i(t)$ stands for the jamming beamforming vector of EV. 

By using the self interference cancellation techniques\cite{5}, the EV is able to mitigate the residual SI to a zero-mean circularly symmetric complex Gaussian (CSCG) AWGN, denoted by $n_1 \sim \mathcal{CN}(0,\sigma^2_I)$. Consequently, the received signal at EV can be expressed as
\begin{equation}
    \begin{split}
        y_{i,t}^\mathcal{E} = &(\sum \nolimits _{k=1}^K g^H_{ke,t}{\Phi }_k(t)G^H_{ak,t}+h^H_{ae,t})w_i(t)s_{i,t}+\\&\sum \nolimits _{j\ne i}(\sum \nolimits _{k=1}^K g^H_{ke,t}{\Phi}_k(t)G^H_{ak,t}+h^H_{ae,t})w_j(t)s_{j,t}\\&+n_1+n_0.
    \end{split}
\end{equation}

For simplicity, we let $H_{ai,t}=\sum \nolimits _{k=1}^K g^H_{ki,t}{\Phi }_k(t)G^H_{ak,t}+h^H_{ai,t}$, $H_{ae,t}=\sum \nolimits _{k=1}^K g^H_{ke,t}{\Phi }_k(t)G^H_{ak,t}+h^H_{ae,t}$ and $H_{ei,t}=h^H_{ei,t}$. Then the signal-to-interference-plus noise ratio (SINR) of received signal at the $i$th LR can be calculated by
{$SINR_{i,t}^\mathcal{L}\hspace{-0.5mm}=\hspace{-0.5mm}{|H_{ai,t}w_i(t)|^2}\hspace{-0.5mm}/\hspace{-0.5mm}{(\sum \nolimits _{j\ne i}\hspace{-0.5mm}|H_{ai,t}w_j(t) |^2\hspace{-0.5mm}+\hspace{-0.5mm}\sum \nolimits _{j=1}^L \hspace{-0.5mm}|H_{ei,t}f_j(t)|^2\hspace{-0.5mm}+N_0)}$}, where $N_0$ is the power of AWGN at $i$th LR. Thus, the achievable rate of $i$th LR can be formulated as $R_{i}^{\mathcal{L}}(t)=\log(1+SINR_{i,t}^{\mathcal{L}})$. Similarly, the SINR of the $i$th LR's signal at the EV can be calculated by
$SINR_{i,t}^\mathcal{E}={|H_{ae,t}w_i(t)|^2}/{(\sum \nolimits _{j\ne i}|H_{ae,t}w_j(t) |^2+N_1+N_0)}$, where $N_1$ denotes the mitigated SI power at the EV. Therefore, the achievable rate of EV can be formulated as $R_{i}^{\mathcal{E}}(t)=\log(1+SINR_{i,t}^\mathcal{E}).$ Then, according to Wyner's definition of PLS\cite{20}, the secrecy rate of $i$th LR can be formulated as 
\begin{equation}
    R_i^{sec}(t)=[R_i^{\mathcal{L}}(t)-R_i^{\mathcal{E}}(t)]^+,
\end{equation}where function $[x]^+=\max\{x,0\}$.

\subsection{Problem formulation}
In this subsection, we formulate three respective utility functions for the three parties, along with their respective optimization problems (i.e., LPs, TIRSs, EV) in PLS under system dynamics. 

For LPs, to improve their long-term secrecy performance, while reducing their power consumptions, in each time slot $t$, they need to determine i) LT's transmit beamforming vectors $w_i(t), \forall i \in \mathcal{L}$; ii) the payment $\mu^{\mathcal{R}}_{\mathcal{L}}$ for attracting TIRSs' help. Then LPs' utility function in time slot $t$ can be expressed as
\begin{equation}
    \begin{split}
        U^{\mathcal{L}}(t)= &\sum\nolimits_{i=1}^{L}R_i^{sec}(t) - c_1(t)\mu^{\mathcal{R}}_{\mathcal{L}}(t) \\&- \rho (E^{\mathcal{L}}(t) + c_1(t) E^{\mathcal{R}}(t)),
    \end{split}
\end{equation}
where $c_1(t) = 1$ or $0$ denotes whether TIRSs choose to assist LPs or not, $\rho$ is the unit power cost, $E^{\mathcal{L}}(t) = \sum \nolimits^L_{i=1}\xi^{\mathcal{L}}{w_i(t)}^Hw_i(t)+P_B+\sum \nolimits^L_{i=1}P_i$ and $E^{\mathcal{R}}(t) = \sum \nolimits^K_{k=1}NP^{\mathcal{R}}$ denote the total power consumption of LPs and TIRSs, respectively, where $\xi^{\mathcal{L}}$ is the amplifier coefficient of the LT, $P_B$ denotes the circuit power of the LT's transmission, $P_i$ denotes the circuit power consumption of the $i$th LR, and $P^{\mathcal{R}}$ is the power consumption of each reflecting element in TIRSs. With LPs' strategies denoted as $\pi^{\mathcal{L}}=\{w_i(t),\mu^{\mathcal{R}}_{\mathcal{L}}(t)\}_{\forall i, \forall t}$, the long-term optimization problem for LPs can be formulated as
\begin{align}
    [\rm {\mathcal{LP}}]:&\argmax_{\pi^{\mathcal{L}}}\lim_{T \to \infty}\frac{1}{T}\sum\nolimits_{t=0}^{T}U^{\mathcal{L}}(t)
        \\ \quad ~~s.t.~\;&R_i^{sec}(t)\ge R_{min}^{sec},\forall i \in \mathcal{L},\tag{5a}
        \\ \hphantom {\quad ~~s.t.~} &\|w_i(t)\|^2\le P_{max}^{L},\forall i \in \mathcal{L},\tag{5b}
        \\ \hphantom {\quad ~~s.t.~} &0 \le \mu^{\mathcal{R}}_{\mathcal{L}}(t)\le \sum\nolimits_{i=1}^{L}R_i^{sec}(t), \tag{5c}
\end{align}
where $R_{min}^{sec}$ is the minimum required secrecy rate, and constraint (5b) means that the transmit power of LT for each LRs cannot exceed the maximum transmit power $P^\mathcal{L}_{max}$.

For EV, to improve its long-term eavesdropping performance, while reducing its power consumption from eavesdropping and jamming. In each time slot $t$, they need to determine i) the jamming beamforming vectors $f_i(t), \forall i \in \mathcal{L}$; ii) the payment $\mu^{\mathcal{R}}_{\mathcal{E}}$ for attracting TIRSs' help. The utility function for the EV in time slot $t$ can be expressed as
{\begin{equation}
    \begin{split}
        U^{\mathcal{E}}(t)= &-\sum\nolimits_{i=1}^{L}R_i^{sec}(t) - c_2(t)\mu^{\mathcal{R}}_{\mathcal{E}}(t) \\ &- \rho (E^{\mathcal{E}}(t) + c_2(t) E^{\mathcal{R}}(t)),
    \end{split}
\end{equation}}where $c_2(t)=1$ or $0$ denotes whether TIRSs choose to assist EV or not, and $E^{\mathcal{E}}(t) = \sum \nolimits^\mathcal{L}_{i=1}\xi^{\mathcal{E}}{f_i(t)}^Hf_i(t)+P^\mathcal{E}$ denotes the total power consumption of EV, where $\xi^{\mathcal{E}}$ is the amplifier coefficient of the EV, $P^\mathcal{E}$ denotes the circuit power of EV's eavesdropping. With EV's strategies denoted as $\pi^{\mathcal{E}}=\{f_i(t),\mu^{\mathcal{R}}_{\mathcal{E}}(t)\}_{\forall i,\forall t}$, the long-term optimization problem for EV can be formulated as
\begin{align}
    [\rm {\mathcal{EP}}]:&\argmax_{\pi^{\mathcal{E}}}\lim_{T \to \infty}\frac{1}{T}\sum\nolimits _{t \in [0,T)}U^{\mathcal{E}}(t)
    \\ \quad ~~s.t.~\; &\sum\nolimits_{i=1}^\mathcal{L}\|f_i(t)\|^2\le P_{max}^{\mathcal{E}}, \tag{7a}
    \\ \hphantom {\quad ~~s.t.~} &0 \le \mu^{\mathcal{R}}_{\mathcal{E}}(t)\le \sum\nolimits_{i=1}^{L}R_i^{sec}(t), \tag{7b} 
\end{align}
where the constraint (7a) means that the jamming power of EV for all LR in each time slot cannot exceed the maximum jamming power $P^\mathcal{E}_{max}$.

For TIRSs, in each time slot $t$, they need to decide i) the phase-shifting matrix $\Phi_k,\forall k \in \mathcal{K}$; ii) the alliance selection, denoted as $c_1(t)$ and $c_2(t)$, to maximize their total reward. The utility function for TIRSs can be expressed as
{\begin{equation} 
    \begin{split} 
        U^{\mathcal{R}}(t)= &c_1(t)\mu^{\mathcal{R}}_{\mathcal{L}}(t)+c_2(t)\mu^{\mathcal{R}}_{\mathcal{E}}(t)\\& - C_{conf}\mathcal{F}(t)U^{\mathcal{R}}(t-1),
    \end{split} 
\end{equation}}where $C_{conf}$ denotes the punishment coefficient of the coalition change, and $\mathcal{F}(t)=c_1(t) \oplus c_1(t-1)$. With TIRSs' strategy denotes as $\pi^{\mathcal{R}}=\{\Phi_k,c_1(t),c_2(t)\}_{\forall k,\forall t}$, the long-term optimization problem for TIRSs can be formulated as
\begin{align}
    [ \rm {\mathcal{IP}}]:&\argmax_{\pi^{\mathcal{R}}}\lim_{T \to \infty}\frac{1}{T}\sum\nolimits _{t \in [0,T)}U^{\mathcal{R}}(t)
    \\  \quad ~~s.t.~\; &0\le \theta_{k,n}(t) \le 2\pi, \forall k \in \mathcal{K}, 1 \hspace {-1mm}\le \hspace {-1mm} n\hspace {-1mm} \le\hspace {-1mm} N, \tag{9a}
    \\  \hphantom {\quad ~~s.t.~} &c_1 + c_2 = 1 \tag{9b}.
\end{align}

\section{Game Analyses And DRL-based Approach}
\subsection{Formulation and Analysis on RCFG}
To better describe the dynamic coalition formation among LPs, EV, and TIRSs, we particularly introduce a RCFG, defined as $\mathcal{G}=\{\mathcal{N}, \Delta , \mathcal{U}, \Pi^N, \Pi^C\}$, where $\mathcal{N}=\{\mathcal{L},\mathcal{E},\mathcal{R}\}$ is the set of players in PLS, $\Delta=\{\{\mathcal{L}\},\{\mathcal{E}\},\{\mathcal{L},\mathcal{R}\},\{\mathcal{E},\mathcal{R}\}\}$ represents the set of all possible coalitions in this game, $\mathcal{U}=\{U^{\mathcal{L}}, U^{\mathcal{R}}, U^{\mathcal{E}}\}$ respectively denote their utility functions, as defined in (4), (6) and (8), $\Pi^N=\{\pi_i\}_{\forall i \in \mathcal{N}}$ denotes the non-coalitional strategies of each player $i \in \mathcal{N}$, and $\Pi^C=\{\psi_i\}_{\forall i \in \mathcal{N}}$ denotes the coalitional strategies of each player $i \in \mathcal{N}$, which is equivalent to $c_1$ and $c_2$ in problem formulation. The utility function of each coalition $\mathcal{S}_i \in \Delta$ is defined as
{
\small    
\begin{equation}
U^{\mathcal{S}_i}(t)\hspace{-1mm}=\hspace{-1mm}
\begin{cases}
\sum\nolimits_{i=1}^{L}R_i^{sec}(t) - \sum_{j\in{\mathcal{S}_i}} \rho E^i(t), & \text{if } \mathcal{L} \in \mathcal{S}_i,\\
- \sum\nolimits_{i=1}^{L}R_i^{sec}(t) - \sum_{j\in{\mathcal{S}_i}}\rho E^i(t) ,  &\text{if } \mathcal{E} \in \mathcal{S}_i,\\
\end{cases}
\end{equation}}

In order to guarantee the fairness of incentive payment, $\mu^{\mathcal{R}}_{\mathcal{L}}$ and $\mu^{\mathcal{R}}_{\mathcal{E}}$ is set to be the Shapley value 
of TIRSs' assisting LPs or EV, respectively, and such value can be expressed as
{\small \begin{align}
		\phi^{i}_{\mathcal{S}_j}(t) \hspace{-0.8mm}=\hspace{-0.8mm} \sum\nolimits_{s \in \mathcal{S}_j\setminus i} \hspace{-0.8mm}({|s|!(|\mathcal{S}_j|\hspace{-0.8mm}-\hspace{-0.8mm}|s|\hspace{-0.8mm}-\hspace{-0.8mm}1)!}(U^{s\cup i}(t)\hspace{-0.8mm}-\hspace{-0.8mm}U^s(t)))/{|\mathcal{S}_j|!}, \notag
\end{align}}which measures the contribution of player $i, \forall i \in \mathcal{N}$ in improving $\mathcal{S}_j$'s utility. Then, the payment of TIRSs' assisting LPs or EV, i.e., $\mu^{\mathcal{R}}_{\mathcal{L}}(t)$ and $\mu^{\mathcal{R}}_{\mathcal{E}}(t)$, can be redefined as 
\begin{align}
	\mu^{\mathcal{R}}_i(t)\hspace{-0.8mm}=\hspace{-0.8mm}({U^{\{i,\mathcal{R}\}}(t)\hspace{-0.8mm}+\hspace{-0.8mm}U^{\{\mathcal{R}\}}(t)\hspace{-0.8mm}-\hspace{-0.8mm}U^{\{i\}}(t)})/{2}, \forall i \in \{\mathcal{L},\mathcal{E}\}. \notag
\end{align}

Since each party has different preferences in allying with each other for a better utility, the players in $\mathcal{N}$ own their respective time-varying preference orders for joining in different potential coalitions, which are defined as follows: 

\textbf{\textit{Definition 1}} (\textit{Preference order}): Any party $i \in \mathcal{N}$ prefers to join coalition $\mathcal{S}_a \hspace{-0.4mm}\in \hspace{-0.4mm}\Delta$ than $\mathcal{S}_b \hspace{-0.5mm}\in \hspace{-0.5mm}\Delta$ if $i$ can obtain more utility in $\mathcal{S}_a$ and its member $\forall k \hspace{-0.5mm}\in \hspace{-0.5mm} \mathcal{S}_a \hspace{-0.5mm} \setminus \hspace{-0.5mm} i$ welcome $i$'s joining, i.e.,
\begin{align}
	\mathcal{S}_a \hspace{-0.1mm}\succ_i^t \hspace{-0.1mm}\mathcal{S}_b \hspace{-0.1mm}\Leftrightarrow \hspace{-0.1mm} U^i_{\mathcal{S}_a}(t) \hspace{-0.1mm}>\hspace{-0.1mm} U^i_{\mathcal{S}_b}(t) \text{ and }  U^k_{\mathcal{S}_a}(t) > U^k_{\mathcal{S}_a \setminus i}(t), 
\end{align}

Although the above preference order may guide each party (i.e., LPs, EV, and TIRSs) to select the most beneficial coalition, since this order has to be applied to all three parties, any coalition partitions formed without considering all allies' selfishness within the coalition is unstable. For example, if $\mathcal{R}$ selfishly choose to form a coalition with $\mathcal{L}$, which will decrease the utility of $\mathcal{L}$, such coalition partition does not satisfy the inherent selfishness of all members in the coalition. Therefore, we need to consider the hedonic nature of the three parties in their coalition formation, i.e., i) the utility of each player depends on not only itself but also its allies; and ii) The coalitions form as a result of the preferences of
the players over their possible coalitions' set, in which the switch operation needs to be considered, which is defined as follows:

\textbf{\textit{Definition 2}} (\textit{Switch operation}): Any party $i \in \mathcal{S}_a$ chooses to leave its current coalition $\mathcal{S}_a\in \hspace{-0.4mm}\Delta$ and join another coalition $\mathcal{S}_b\in \hspace{-0.4mm}\Delta$ when the following condition holds.
{\normalsize \begin{align} \label{eq:switch}
	\mathcal{S}_a \hspace{-0.6mm}\rightarrow_i^t \hspace{-0.4mm}\mathcal{S}_b \hspace{-0.4mm}\Leftrightarrow  \hspace{-0.4mm}\mathcal{S}_b \hspace{-0.4mm}\cup \hspace{-0.4mm}\{i\} \hspace{-0.8mm}\succ_i^t \hspace{-0.4mm}\mathcal{S}_a, \hspace{-0.8mm}\text{ and }\hspace{-0.4mm} \mathcal{S}_b \hspace{-0.4mm}\cup \hspace{-0.4mm}\{i\} \hspace{-0.8mm}\succ_k^t \hspace{-0.4mm}\mathcal{S}_b, \hspace{-0.4mm} \forall k \hspace{-0.4mm}\in \hspace{-0.4mm}\mathcal{S}_b.
\end{align}}


In order to achieve the long-term optimal strategies of all three parties while solving the optimization problems $[{\mathcal{LP}}]$, $[{\mathcal{EP}}]$ and $[{\mathcal{IP}}]$ under their potential alliances and confrontations, the existence of Nash equilibrium (NE) of $\mathcal G$ should be rigorously analyzed\cite{30}, which is defined as follows.

{
    \textbf{\textit{Definition 3}} (\textit{Equilibrium of $\mathcal{G}$}): In $\mathcal{G}$, strategy profile $\{\pi^*_i,\psi^*_i\}_{\forall i \in \mathcal{N}}$ is an equilibrium if and only if no player can benefit by unilaterally deviating from $\pi^*_i$ and $\psi^*_i$, i.e.,
    {\small \begin{align}
    	\frac{1}{T}\hspace{-0.4mm}\sum\nolimits_{t=1}^{T} \hspace{-0.4mm}U^i(t) | \pi^*_i,\hspace{-0.4mm}\pi^*_{-i},\hspace{-0.4mm}\psi_i^*,\hspace{-0.4mm}\psi_{-i}^*  \hspace{-0.4mm}\geq \hspace{-0.4mm} \frac{1}{T}\hspace{-0.4mm}\sum\nolimits_{t=1}^{T} \hspace{-0.4mm}U^i(t) | \pi_i,\hspace{-0.4mm}\pi^*_{-i},\hspace{-0.4mm}\psi_i,\hspace{-0.4mm}\psi_{-i}^* \notag
    \end{align}}where $-i$ denotes other players except $i$. 
}

From Definition 3, the NE of RCFG $\mathcal{G}$ strictly depends on the equilibrium coalition strategies of each party, i.e., $\psi^*_i$, which necessitates the exploration on the stability of coalition partition $\Psi(t)$ in each time slot\cite{10286340}. The stable coalition partition is defined as follows:

\textbf{\textit{Definition 4}} (\textit{Stable coalition partition in each time slot}): A coalition partition $\Psi(t) =\{\mathcal{S}_1,...,\mathcal{S}_{|\Psi|}\}$ is individually stable in each time slot if there is no player $i \in \mathcal{N}$ can benefit by changing its current coalition unilaterally, i.e.,
\begin{equation}
    \begin{split}
      U^i_{(\psi_i^*,\psi_{-i}^*)}(t) \geq  U^i_{(\psi_i,\psi_{-i}^*)}(t), \forall i \in \mathcal{N},  \psi_i \neq \psi_i^*.
    \end{split}
\end{equation}

\textbf{\textit{Theorem 1}} (\textit{Existence of the stable coalition partition in each time slot}): By iteratively adopting the switch operations among the three parties (i.e., LPs, EV and TIRSs), the coalition partition $\Psi(t)$ among them can finally converge to a stable coalition partition $\Psi(t)^*$ in each time slot $t$.{
\begin{proof}
    This proof is omitted due to the page limitation.
\end{proof}
}

\textbf{\textit{Theorem 2}} (\textit{Existence of NE in RCFG $\mathcal{G}$}): Given the optimal non-coalitional strategies {\small$\Pi^{N*} = \{\{w_i^*\}_{\forall i}, \{f_i^*\}_{\forall i},\{\Phi_k^*\}_{\forall k}\}_{\forall t}$}, in the proposed RCFG $\mathcal{G}$, there exists at least one NE.
{
\begin{proof}
    This proof is omitted due to the page limitation.
\end{proof}
}

\subsection{Switch Operations-based Coalition Selection with DRL-based Solution for RCFG $\mathcal G$}
According to Theorem 2, in order to achieve the NE of RCFG $\mathcal{G}$, $\Pi^N$ should be optimized first given the specific coalition partition $\Psi$ to achieve long-term optima $\Pi^{N*}$. Then $\Pi^C$ will be optimized by performing the switch operations among the three parties. Owing to the face that, given the specific $\Psi$, within a coalition $\mathcal{S}_i$ in $\Psi$, the current system state (e.g., channel gains) only depends on that in the previous time slot, and the non-coalitional strategies of all members in $\Psi$, the decision-making process of all members in a coalition can be formulated as an MDP, which can be expressed as follows.
{
\footnotesize
\begin{algorithm}[!t]
    \footnotesize
    \caption{Switch Operations-based Coalition Selection with DRL-based Solution for RCFG $\mathcal G$}
    \label{alg:algorithm1}
    \SetKwFunction{Union}{Union}\SetKwFunction{FindCompress}{FindCompress}
    \KwIn{$P_{max}^{\mathcal{L}}$, $P_{max}^{\mathcal{E}}$, $R_{min}^{sec}$, $C_{conf}$}
    \KwOut{${\pi^{\mathcal{L}}}^*, {\pi^{\mathcal{R}}}^*, {\pi^{\mathcal{E}}}^*, \Psi^*$}
    \BlankLine
    \textit{Offline Pretraining stage:}\\
    \For{all possible coalition partitions $\Psi_i$}{
      Initialize $\varphi$ and $\theta$ of $\{At^{\mathcal{S}_i}\}_{\forall \mathcal{S}_i \in \Psi_i}$\;
      \For{each episode $=1,2,\dots, N^{epi}$}{
         Observe initial system states for each coalition\;
         \For{each step $\tau = 0,1,2,\dots, \varGamma$}{
            $\{At^{\mathcal{S}_i}\}_{\forall \mathcal{S}_i \in \Psi_i}$ generates $a^{\mathcal{S}_i}_\tau$ based on $s^{\mathcal{S}_i}_\tau$ and observe $s^{\mathcal{S}_i}_{\tau+1}$ and $r^{\mathcal{S}_i}_\tau$\;
        store {\footnotesize$(s^{\mathcal{S}_i}_\tau,a^{\mathcal{S}_i}_\tau,r^{\mathcal{S}_i}_\tau,s^{\mathcal{S}_i}_{\tau+1})$} in the replay buffer\;
         }
         \If{the replay buffer is full}{
            update the $\varphi$ and $\theta$ of $\{At^{\mathcal{S}_i}\}_{\forall \mathcal{S}_i \in \Psi_i}$\;
         }
      }
    }
    \BlankLine
    \textit{Online implementation stage:}\\
    \For{time slot $t=1,2,\dots$}{
      \Repeat{$\Psi(t)$ converges to $\Psi(t)^*$}{
        TIRSs select a random coalition $\mathcal{S}_j$ to join, reforming the $\Psi(t)$ into $\Psi(t)^{'}=\{\mathcal{S}_a, \mathcal{S}_b\}$\;
        \Repeat{{\footnotesize$U^{\mathcal{L}}_{\mathcal{S}_a^*}(t), U^{\mathcal{E}}_{\mathcal{S}_b^*}(t), U^{\mathcal{R}}_{\mathcal{S}_j^*}(t) $} converge, or the maximum iteration number is reached}{
            $\{At^{\mathcal{S}_i}\}_{\forall \mathcal{S}_i \in \Psi(t)^{'}}$ generates $a^{\mathcal{S}_i}_\tau(t)$ based on $s^{\mathcal{S}_i}_\tau(t)$ and observe $s^{\mathcal{S}_i}_{\tau+1}$ and $r^{\mathcal{S}_i}_\tau$\;
            Calculate the expected utility for all players $U^{\mathcal{L}}_{\mathcal{S}_a^*}(t), U^{\mathcal{E}}_{\mathcal{S}_b^*}(t), U^{\mathcal{R}}_{\mathcal{S}_j^*}(t)$\;
        }
        \For{each player $i \in \mathcal{N}$}{
            perform the switch operation\;
        }
      }
    }\vspace{-0.5em}
\end{algorithm}
}

\textbf{MDP for all members in one coalition in PLS}: For each coalition $\mathcal{S}_i \in \Delta$ in PLS, its corresponding MDP is expressed as $\mathcal{M}_{\mathcal{S}_i}$ = $\{s^{\mathcal{S}_i}(t), a^{\mathcal{S}_i}(t), \Xi^{\mathcal{S}_i}(t), r^{\mathcal{S}_i}(t)\}$.

1) \textit{Environment State for Each Coalition in PLS:} For each coalition $\mathcal{S}_i$ in time slot $t$, its environment state can be expressed as $s^{\mathcal{S}_i}(t)=\{H(t), R^{\mathcal{L}}(t),R^{\mathcal{E}}(t),R^{sec}(t)\}$, where $H(t)$ denotes the equivalent channel gains of all links in the system, $R^{\mathcal{L}}(t)=\{R^{\mathcal{L}}_i(t)\}_{\forall i}$ if LPs is in the coalition $\mathcal{S}_i$, otherwise $R^{\mathcal{L}}(t)=\varnothing$, $R^{sec}(t)=\{R^{sec}_i(t)\}_{\forall i}$, $R^{\mathcal{E}}(t)=\{R^{\mathcal{E}}_i(t)\}_{\forall i}$ if EV is in the coalition $\mathcal{S}_i$, otherwise $R^{\mathcal{E}}(t)=\varnothing$;

2) \textit{Action for Each Coalition in PLS:} In time slot $t$, each coalition $\mathcal{S}_i$'s action is denoted as $a^{\mathcal{S}_i}(t)\hspace{-1.1mm}=\hspace{-1.1mm}\{a^{j}(t)\}_{\forall j \in {\mathcal{S}_i}}$, where $a^{\mathcal{L}}(t)\hspace{-1.1mm}=\hspace{-1.1mm}\{w_i(t)\}_{\forall i},a^{\mathcal{E}}(t)\hspace{-1.1mm}=\hspace{-1.1mm}\{f_i(t)\}_{\forall i}, a^{\mathcal{R}}(t)\hspace{-1.1mm}=\hspace{-1.1mm}\{\Phi_k(t)\}_{\forall k}$ are exactly the same as the non-coalitional strategies of each player $i \in \mathcal{N}$. We assume that each coalition $\mathcal{S}_i$'s decision-making is delegated to distinct central controllers \cite{12};

3) \textit{State Transition Probabilities of Each Coalition in PLS:} The state transition probability from $s^{\mathcal{S}_i}(t)$ to ${s^{\mathcal{S}_i}(t)}^{'}$ by taking $a^{\mathcal{S}_i}(t)$ is expressed as $\Xi^{\mathcal{S}_i}(t)=Pr({s^{\mathcal{S}_i}(t)}^{'}|s^{\mathcal{S}_i}(t),a^{\mathcal{S}_i}(t))$;

4) \textit{Reward of Each Coalition in PLS:} In time slot t, the immediate reward of coalition $\mathcal{S}_i$ is denoted as $r^{\mathcal{S}_i}(t)= U^{S_i}(t) - \sum\nolimits_{j \in \mathcal{L}}\eta  p^{sec}_j$, where $p^{sec}_j=1$    if $\mathcal{L} \in \mathcal{S}_i$ and $R^{sec}_j(t) < R^{sec}_{min},\forall j \in \mathcal{L}$ hold, otherwise $p^{sec}_j=0$. The coefficient $\eta$ is positive constant of the second part, which is used to balance the utility and the security requirement. 

Due to the inherent interdependence in three parties' decision-making, the MDPs for each coalition in PLS are tightly coupled. Furthermore, due to system dynamics and multi-dimensional decisions of each party, the state space and action space $\mathcal{M}_{\mathcal{S}_i},\forall \mathcal{S}_i \in \Delta$ become relatively large. To address this challenge, we introduce a DRL-based algorithm for the three parties' optimal non-coalitional strategies, which includes an actor-critic (AC) scheme for action generation/evaluation, and a proximal policy optimization (PPO)-based network updating for optimal strategies. For each coalition $\mathcal{S}_i$, AC includes i) a critic network with parameter $\varphi$ to estimate $\mathcal{S}_i$'s state value $V^{\mathcal{S}_i}_{\varphi}(s^{\mathcal{S}_i}(t))$; ii) an actor network with parameter $\theta$ to generate action $a^{\mathcal{S}_i}(t)$ based on the state $s^{\mathcal{S}_i}(t)$. 

In order to solve the equilibrium of $\mathcal{G}$, i.e., $\{\Pi^{C*},\Pi^{N*}\}$, we introduce a switch operations-based coalition selection along with DRL-based algorithm, which consists of an offline pretraining stage and an online implementation stage as follows. 

\textbf{Offline pretraining stage (solution for $\Pi^{N*}$)}: We first train the agents of each coalition under all possible coalition partitions. The decision-making of any coalition $\mathcal{S}_i \in \Psi_i$ is assigned to agent $At^{\mathcal{S}_i}$ with an actor network $\theta^e$ and a critic network $\varphi^e$. In every training episode, all agents interact within same time slots, performing decision-making process for each party. Specifically, in each training step $\tau$, each agent $At^{\mathcal{S}_i}, \forall \mathcal{S}_i \in \Psi_i$ observes the current state $s^{\mathcal{S}_i}_\tau$ and generates the action $a^{\mathcal{S}_i}_\tau$ to get reward $r^{\mathcal{S}_i}_\tau$. Then the current state changes to $s^{\mathcal{S}_i}_{\tau+1}$, and tuple $(s^{\mathcal{S}_i}_\tau, a^{\mathcal{S}_i}_\tau, r^{\mathcal{S}_i}_\tau, s^{\mathcal{S}_i}_{\tau+1})$ is stored in the replay buffer for network updating. When the replay buffer is full, using $e$ to indicate any agent $At^{\mathcal{S}_i}, \forall \mathcal{S}_i\in \Psi_i$, the network updating process includes i) calculating $e$'s rewards-to-go $J^e_\tau\hspace{-0.5mm}=\hspace{-0.5mm}\sum_{\tau'=\tau}^{\varGamma}\gamma^{\tau'\hspace{-0.5mm}-\tau}r^e_{\tau'}$; ii) calculating $e$'s advantage function $A^e_\tau\hspace{-0.5mm}=\hspace{-0.5mm}J^e_\tau \hspace{-0.5mm}-\hspace{-0.5mm} V^e_{\varphi}(s^e_\tau)$; iii) calculating the loss function of $\theta^e$, i.e., $L^{CLIP}(\theta^e)\hspace{-0.5mm}=\hspace{-0.5mm}\sum\nolimits^\varGamma_{\tau=0}\text{min}(\frac{\pi_{\theta^e} (a_\tau, s_\tau)}{\pi_{\theta^e_{old}} (a_\tau, s_\tau)}A^e_\tau,\hspace{-0.3mm} \textit{clip}(\frac{\pi_{\theta^e} (a_\tau, s_\tau)}{\pi_{\theta^e_{old}} (a_\tau, s_\tau)},\hspace{-0.5mm} 1-\epsilon ,\hspace{-0.5mm}1+\epsilon) A^e_\tau)$, where 
$\pi_{\theta^e}(a_\tau,\hspace{-0.5mm} s_\tau)$ is the probability of $\theta^e$'s choosing action $a_\tau$ at state $s_\tau$, $\theta^e_{old}$ is the original parameter of $theta^e$ before updating; and iv) calculating the loss function of $\varphi^e$, i.e.,  $L^{VF}(\varphi^e)\hspace{-0.5mm}=\hspace{-0.5mm}\sum\nolimits^{\varGamma}_{\tau=0}(V^e_{\varphi^e}(s^e_\tau)\hspace{-0.5mm}-\hspace{-0.5mm}J^e_\tau)^2$. Then, $\theta^e$ and $\varphi^e$ are updated by minimizing their corresponding loss functions via a random gradient descent. After $N^{epi}$ episodes, the trained actor networks $\{\theta^e\}_{\forall e}$ are equivalent to long-term optimal non-coalitional strategies of all coalitions.

\textbf{Online Implementation stage (solution for $\Pi^{C*}$)}: After obtaining the optimal non-coalitional strategies ${\Pi^N}^*$ trained in the previous stage, the three players iteratively adopt the switch operation in (\ref{eq:switch}) to form the stable coalition partition $\Psi^*(t)$ in each time slot to obtain the optimal coalition strategies ${\Pi^C}^*$. To be more specific, in each iteration, i) TIRSs first select a random coalition $\mathcal{S}_j$ to join, reforming the $\Psi(t)$ into $\{\mathcal{S}_a, \mathcal{S}_b\}$; ii) then the corresponding agents of coalition $\mathcal{S}_a$ and $\mathcal{S}_b$ iteratively generate their actions according to current state $s^{\mathcal{S}_a}(t)$ and $s^{\mathcal{S}_b}(t)$, and calculate their utilities $U^{\mathcal{L}}_{\mathcal{S}_a}(t), U^{\mathcal{E}}_{\mathcal{S}_b}(t), U^{\mathcal{R}}_{\mathcal{S}_j}(t)$, until their actions keep unchanged; and iii) finally each player iteratively adopt the switch operations until the coalition partition converge to stable coalition partition $\Psi^*(t)$. This process repeats in each time slot. Algorithm \ref{alg:algorithm1} summarizes all detailed steps of the proposed framework.
\begin{figure}[!t]
	\centering
	\includegraphics[width=0.3\textwidth]{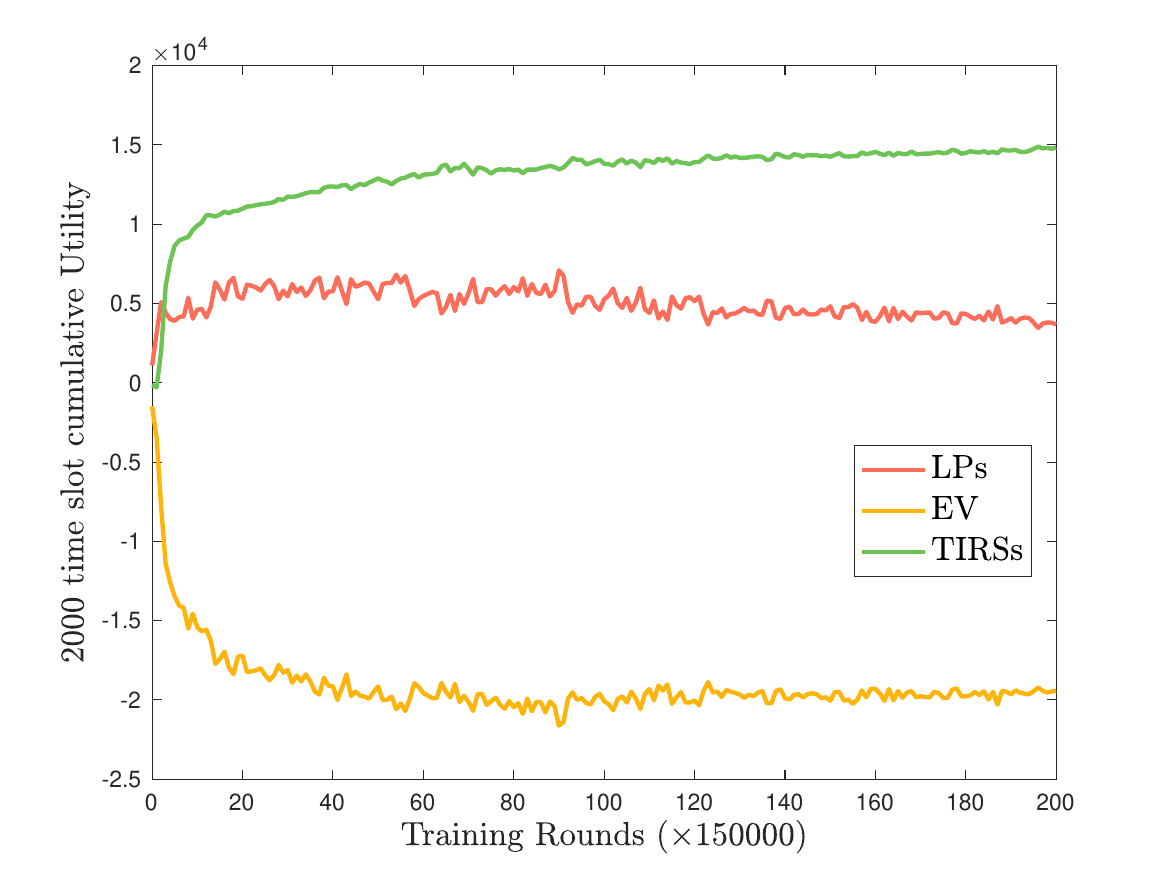}\\   
	\caption{Convergence of the proposed algorithm.}\label{pic:exp1}
	\vspace{-1.6em}   
\end{figure}

\begin{figure*} [t]
	\centering
	\subfloat[]{
		\includegraphics[width=0.3\linewidth]{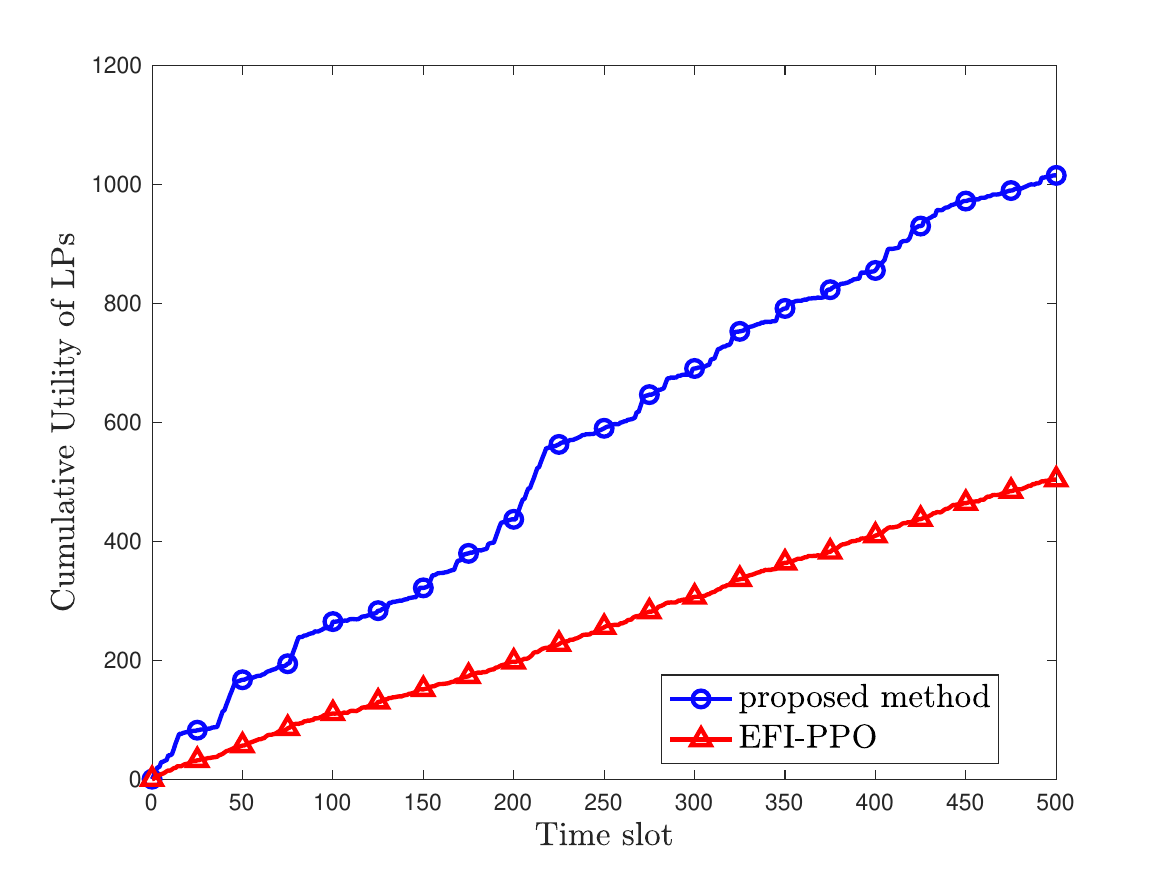}}
        \label{pic:exp2:a}
	\subfloat[]{
		\includegraphics[width=0.3\linewidth]{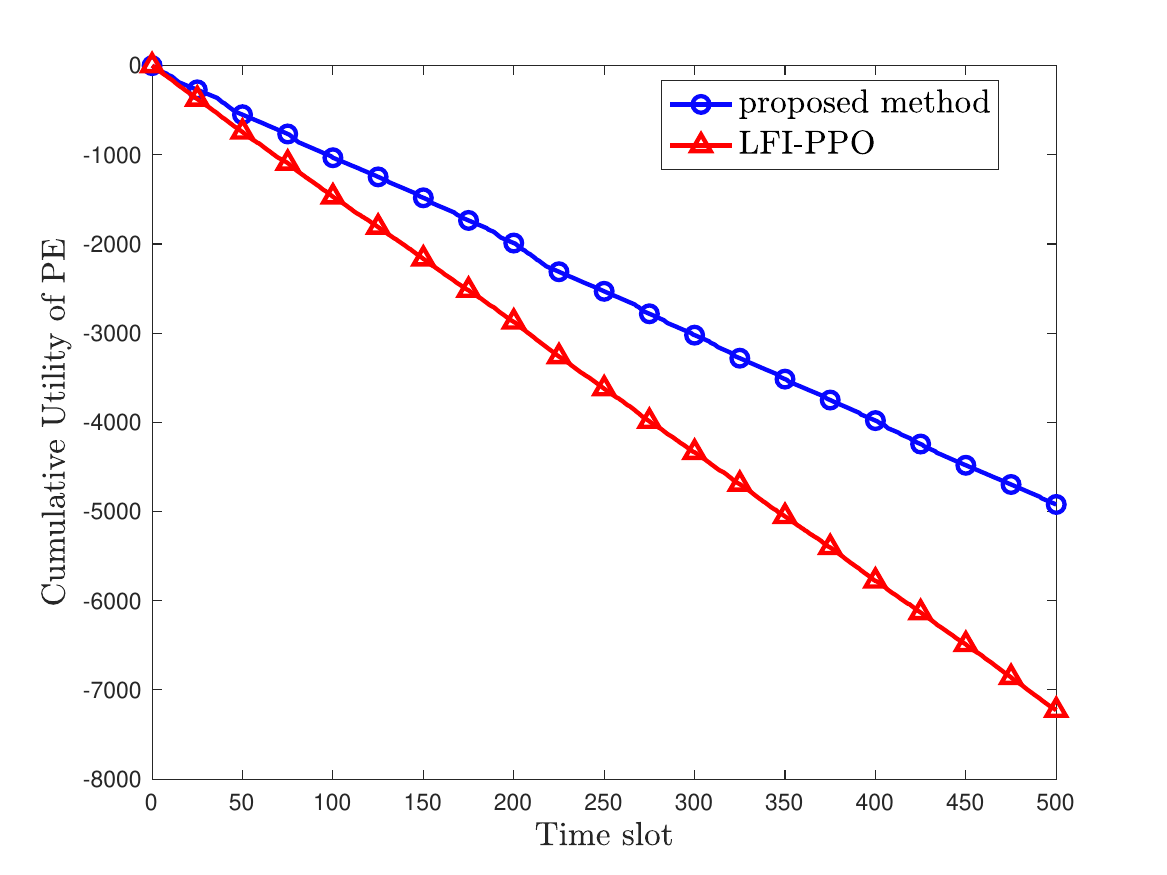}}
        \label{pic:exp2:b}
	\subfloat[]{
		\includegraphics[width=0.3\linewidth]{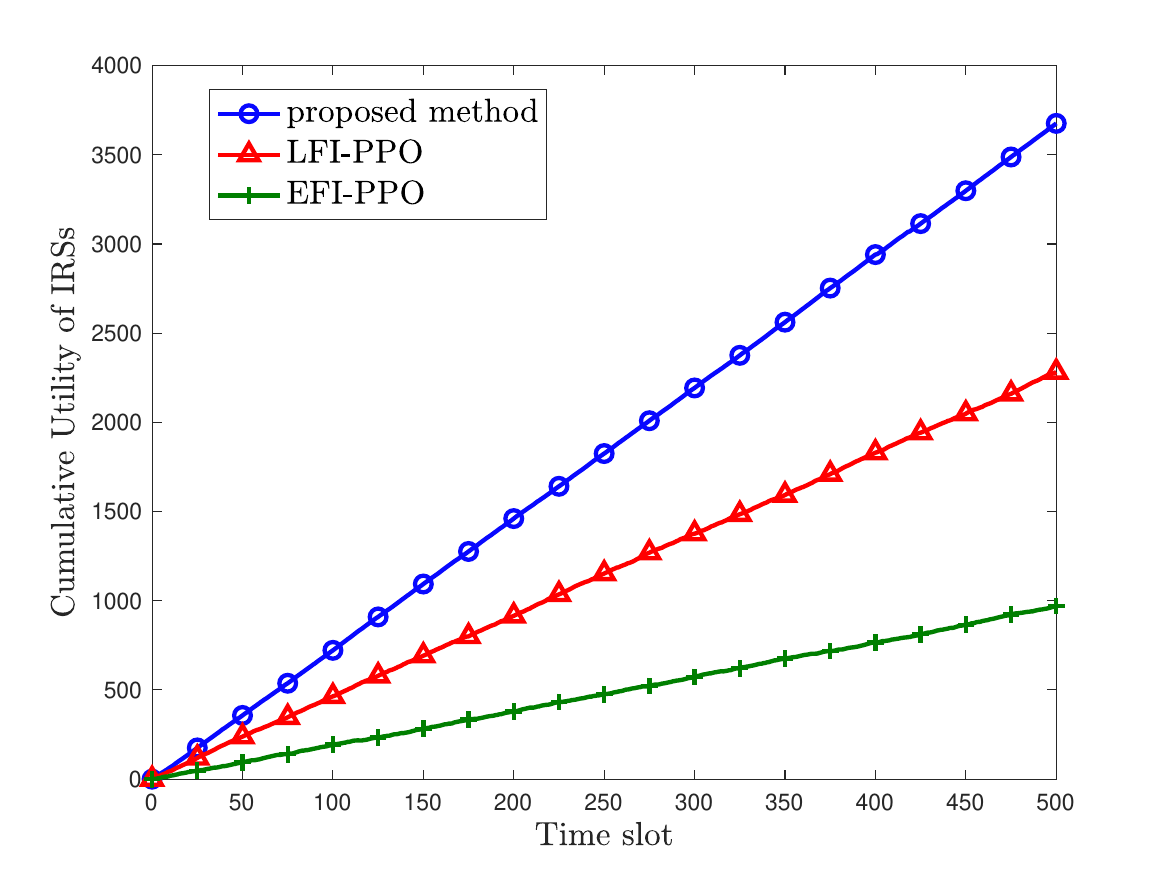}}
        \label{pic:exp2:c}
	\caption{Performance comparison in terms of LPs', EV's and TIRSs' utilities.}
	\label{pic:exp2}
	\vspace{-1.6em}
\end{figure*}
\section{Simulation Results}
We consider a downlink PLS-aware wireless communication system with TIRSs in a $100\hspace{-0.2mm}\times \hspace{-0.2mm}100 \hspace{-0.2mm}\times \hspace{-0.2mm}100 m^3$ Cartesian coordinate area. There exists one LT with $M_a\hspace{-0.5mm}=\hspace{-0.5mm}4$ antennas, one EV with $M_e\hspace{-0.5mm}=\hspace{-0.5mm}4$ antennas, and $L\hspace{-0.5mm}=\hspace{-0.5mm}3$ LRs and $K\hspace{-0.5mm}=\hspace{-0.5mm}2$ IRSs, with $N\hspace{-0.5mm}=\hspace{-0.5mm}16$ reflecting elements each. The direct channels among LT, LRs and EV are blocked by severe obstacles, resulting in much smaller channel gains than that of IRSs' reflecting channels. Following \cite{9}, all channels are modeled as $h_{mn}\hspace{-0.5mm}=\hspace{-0.5mm}\sqrt{L_0d^{-\beta_{mn}}_{mn}}h^*_{mn}$,  in which $L_0 = -30dB$ denotes the reference path loss at 1 meter, $\beta_{mn}$ is the path loss exponent from $m$ to $n$ and is set to $\beta_{ai}\hspace{-0.5mm}=\hspace{-0.5mm}\beta_{ae}\hspace{-0.5mm}=\hspace{-0.5mm}\beta_{ei}\hspace{-0.5mm}=\hspace{-0.5mm}4,\beta_{ak}\hspace{-0.5mm}=\hspace{-0.5mm}\beta_{ke}\hspace{-0.5mm}=\hspace{-0.5mm}\beta_{ki}\hspace{-0.5mm}=\hspace{-0.5mm}2$, $d_{mn}$ stands for the distance from device $m$ to $n$, and $h^*_{mn}\hspace{-0.5mm}=\hspace{-0.5mm}\sqrt{\frac{K^\prime_{mn}}{K^\prime_{mn}+1}}h^*_{LoS}\hspace{-0.5mm}+\hspace{-0.5mm}\sqrt{\frac{1}{K^\prime_{mn}+1}}h^*_{NLoS}$ is the small-scale fading components with Rician factor $K^\prime_{ai}\hspace{-0.5mm}=\hspace{-0.5mm}K^\prime_{ae}\hspace{-0.5mm}=\hspace{-0.5mm}1,K^\prime_{ak}\hspace{-0.5mm}=\hspace{-0.5mm}K^\prime_{ki}\hspace{-0.5mm}=\hspace{-0.5mm}K^\prime_{ke}\hspace{-0.5mm}=\hspace{-0.5mm}10$,  where $h^*_{LoS}$ and $h^*_{NLoS}$ represent the components of the line-of-sight (LoS) and non-line-of-sight channels, respectively, $h^*_{NLoS}$ is i.i.d. complex Gaussian random variable with zero mean and unit variance, and $h^*_{LoS} \hspace{-0.5mm}=\hspace{-0.5mm} a_m(\theta)a_n(\theta)^H$ with $a_m$ and $a_n$ being the array response vectors of the transmitter $m$ and receiver $n$, respectively, which can be expressed as $a_i\hspace{-0.5mm}=\hspace{-0.5mm}[1,..., \exp(j\frac{2\pi}{\lambda}d_i(i-1)\sin(\phi_{LoS_i})cos(\theta_{LoS_i}))]^H,\forall i \in \{m,n\}$, where $d_i\hspace{-0.5mm}=\hspace{-0.5mm}1/2\lambda$ is the inter-antenna spacing of $i$, $\phi_{LoS_i}$ and $\theta_{LoS_i}$ stands for the azimuth and elevation angles of $i$, respectively. 
Furthermore, $\eta\hspace{-0.5mm}=\hspace{-0.5mm}2$, $C_{conf}\hspace{-0.5mm}=\hspace{-0.5mm}0.1$, $P^\mathcal{L}_{max}\hspace{-0.5mm}=\hspace{-0.5mm}40$dBm, $P^\mathcal{E}_{max}\hspace{-0.5mm}=\hspace{-0.5mm}15$dBm, $\xi^{\mathcal{L}}\hspace{-0.5mm}=\hspace{-0.5mm}0.01$, $\xi^{\mathcal{E}}\hspace{-0.5mm}=\hspace{-0.5mm}0.1$, $\rho\hspace{-0.5mm}=\hspace{-0.5mm}0.001$, $N_0\hspace{-0.5mm}=\hspace{-0.5mm}N_1\hspace{-0.5mm}=\hspace{-0.5mm}-174$dBm \cite{10173745}.


Fig. \ref{pic:exp1} examines the convergence of the proposed algorithm. It can be seen that all parties' cumulative utilities exhibit fast convergence, which implies that the strategies of the three parties can converge to the NE under the switch operation-based coalition formation along with PPO-based solution. Moreover, TIRSs' cumulative utility achieved in each training rounds is much more stable than others, this result also give an insight that when NE is achieved, TIRSs receive similar incentive from either LPs or EV, which is in line with the definition of NE that the three parties cannot achieve higher utility by varying their respective strategies.

Fig. \ref{pic:exp2} illustrates the superiority of the proposed algorithm in terms of LPs', EV's and TIRSs' cumulative utilities. For comparison, an EV's friendly phase-shifting policy with PPO-based DRL algorithm (EFI-PPO), in which TIRSs always and only help the EV by adjusting the phase shift of TIRSs to maximize the group utility of the EV coalition\cite{12}, and an LPs' friendly phase-shifting policy with PPO-based DRL algorithm (LFI-PPO), in which TIRSs always and only help the LPs by adjusting the phase shift of TIRSs to maximize the group utility of the LPs coalition\cite{5}. It can be seen that i) in Fig. \ref{pic:exp2}(a), the proposed solution outperforms EFI-PPO in terms of LPs' cumulative utility; ii) in Fig. \ref{pic:exp2}(b), the proposed solution outperforms LFI-PPO in terms of EV's cumulative utility; iii) in Fig. \ref{pic:exp2}(c), the proposed solution outperforms both EFI-PPO and LFI-PPO in terms of TIRSs' cumulative utility. This is because the proposed PPO-based algorithm with RCFG allows TIRSs to dynamically form coalitions with either LPs or EV for higher long-term utilities rather than maintaining fixed relationships in existing studies.

\section{Conclusion}
In this paper, focusing on the scenarios with TIRSs, to model the long-term coalition formations with dynamic decision-making among the LPs, EV and TIRSs in PLS-aware wireless communications, a three-party RCFG is formulated and analyzed. After proving the existence of the NE in proposed RCFG, a switch operations-based coalition selections along with a DRL-based algorithm is proposed to solve the equilibrium strategies for three parties, which maximize their respective long-term performances under dynamic evolutions of coalition partition among them. Simulation results verify the feasibility of the proposed algorithm and demonstrate its superiority over counterparts.

\begin{spacing}{0.96}
	\bibliographystyle{IEEEtran}
	\bibliography{IEEEabrv,ref}
\end{spacing}

\end{document}